\documentclass{article}
\usepackage{graphicx}

\title{Application of entanglement conditions to spin systems}

\author{Hongjun Zheng$^{1}$, Ho Trung Dung$^{1,2}$, and Mark Hillery$^{1}$ \\
$^{(1)}$Department of Physics, Hunter College of CUNY \\ 695 Park Avenue \\ New York, NY 10065 \and
$^{(2)}$ Institute of Physics, Academy of Sciences and Technology \\ 1 Mac Dinh Chi Street,
District 1 \\ Ho Chi Minh City, Vietnam}

\begin{document}
\maketitle

\begin{abstract}
There have been 
numerous studies of entanglement in spin systems.  These have usually focussed
on examining the entanglement between individual spins or determining whether the state of the
system is completely separable.  Here we present conditions that allow us to determine whether 
blocks of spins are entangled.  We show that sometimes these conditions can detect entanglement
better than conditions involving individual spins.  We apply these conditions to study entanglement in
spin wave states, both when there are only a few magnons present and also at finite temperature.
\end{abstract}

\section{Introduction}
The realization that entanglement is a resource for a number of useful tasks in quantum information has led to a tremendous interest in its properties, quantification and in methods by which it can be produced.
One area, which has been fruitful, is the study of entanglement in many-body systems (for a review
see \cite{amico}).

Spin systems, in particular, have received a great deal of attention, and this has led to the formulation
of several conditions for determining whether the state of a spin system is entangled.  For example, 
in an $N$ qubit system, with 
\begin{equation}
J_{l} = \frac{1}{2} \sum_{k=1}^{N} \sigma_{k}^{(l)} ,
\end{equation}
with $l=1,2,3$ and $\sigma_{k}^{(l)}$ being the Pauli matrices for the $k^{\rm th}$ qubit, the state is entangled if \cite{sorensen}
\begin{equation}
\label{spin-ent}
\frac{ (\Delta J_{3})^{2} }{ \langle J_{1}\rangle^{2} + \langle J_{2}\rangle^{2}} < \frac{1}{N} .
\end{equation}
If this inequality is satisfied, the N-qubit state cannot be expressed as
\begin{equation}
\rho = \sum_{j} p_{j} \rho_{j}^{(1)} \otimes \rho_{j}^{(2)} \ldots \otimes \rho_{j}^{(N)} ,
\end{equation}
where the $1\geq p_{j}>0$ sum to one, and $\rho_{j}^{(k)}$ is a density matrix for the $k^{\rm th}$ 
qubit.  A state that can be expressed in this form is known as completely separable.
Further criteria for entanglement in spin systems have been developed \cite{korbicz,toth1,toth2}.
A comprehensive study of entanglement conditions employing quantities that are at most quadratic
in the collective spin operators has been given in \cite{briegel1,briegel2}.

Recently two inequalities have been developed for the detection of entanglement \cite{hillery1}.  
Consider a system consisting of two subsystems, which we shall denote by $a$ and $b$.  The
Hilbert space for the total system is $\mathcal{H} = \mathcal{H}_{a} \otimes \mathcal{H}_{b}$, where
$\mathcal{H}_{a}$ is the Hilbert space for subsystem $a$ and $\mathcal{H}_{b}$ is the Hilbert
space for subsystem $b$.  Let $A$ be an operator on $\mathcal{H}_{a}$ and $B$ be an operator
on $\mathcal{H}_{b}$.  The state of the total system is entangled if either
\begin{equation}
\label{entcongen1}
|\langle AB^{\dagger}\rangle |^{2} > \langle A^{\dagger}AB^{\dagger}B\rangle ,
\end{equation}
or
\begin{equation}
\label{entcongen2}
|\langle AB\rangle |^{2} > \langle A^{\dagger}A\rangle \langle B^{\dagger}B\rangle .
\end{equation}
%
Note that the condition (\ref{entcongen1}) holds if the left-hand
side is replaced by  $|\langle A^{\dagger}B\rangle |^{2}$. In the following we shall 
use the form (\ref{entcongen1}) or 
$|\langle A^{\dagger}B\rangle |^{2} > \langle A^{\dagger}AB^{\dagger}B\rangle$ interchangeably.
These are sufficient conditions for entanglement; if they are not satisfied we cannot say whether the
state is entangled or not.  Some of the implications of these conditions have been explored for the
case that the systems are field modes \cite{hillery1,hillery2}, and in the case that one is a field mode 
and one is a collection of atoms \cite{hillery3}.  Here we would like to explore their implications for
spin systems.  In that case, we have a collections of spins, and our subsystems are two 
non-overlapping subsets of the total set.  We describe each subset by a collective spin, 
$\mathbf{J}_{a}$ for set $a$ and $\mathbf{J}_{b}$ for set $b$.  Let $J_{a-}$ be the angular momentum
lowering operator for set $a$ and $J_{b-}$ be the angular momentum lowering operator for set $b$.  
The corresponding raising operators are $J_{a+}$ and $J_{b+}$, respectively.  Our entanglement
conditions become
\begin{eqnarray}
\label{entcond}
|\langle J_{a-}J_{b+}\rangle |^{2} & > & \langle 
J_{a+}J_{a-}J_{b+}J_{b-} 
\rangle,   
 \nonumber \\
|\langle J_{a-}J_{b-}\rangle |^{2} & > & \langle  J_{a+}J_{a-}\rangle \langle J_{b+}J_{b-}\rangle  .
\end{eqnarray}
These inequalities differ from the ones discussed in the previous paragraph in that they 
detect entanglement between two blocks of spins, and not whether the state is completely separable
or not.  For example, if one is studying entanglement
in a spin chain, one may simply be interested in whether the state is entangled or not, in which case
Eq.\ (\ref{spin-ent}) could be of use.  However, one might instead wish to find out whether two blocks
of spins are entangled, in which case the above equations, with $J_{a}$ and $J_{b}$ being 
collective spin operators for the respective blocks, could be useful.

As was mentioned previously, a large amount of work has been done on entanglement in spin systems, in particular spin chains with different types of interactions between adjacent spins.  
Typically the concurrence of two spins in the ground state 
of the system is found.  This has been done for many variants of the Heisenberg model, both with and without  an applied magnetic field.  We will be interested in applying our conditions to a 
spin system at finite temperature, in particular one with a ferromagnetic Heisenberg
interaction.  Wang and Zanardi showed that in a one-dimensional ring with periodic boundary conditions, and the spins interacting via an isotropic Heisenberg interaction, there is no entanglement between any two spins at any temperature \cite{zanardi}.  This result depends on the $SU(2)$ symmetry 
of the Hamiltonian, and does not appy if there is an applied magnetic field.  Asoudeh and Karimipour look at the case in which there is an applied magnetic field and only the ground state and the state with one flipped spin are populated, and found the concurrence between two spins as a function of their separation \cite{asoudeh}.  The advantage of applying the entanglement conditions in the previous
paragraph is that we can study the entanglement between blocks of spins, and not just the
entanglement between individual spins.

The paper is arranged as follows.  We will initially study some general properties of our entanglement
conditions.  We will first compare the application of our conditions to individual and collective spins.  
It will be shown that the conditions can detect entanglement
between spins in angular momentum intelligent states.  These are states for which the uncertainty
relations for angular momentum operators are satisfied as an equality.  We will then show how the
above conditions can be strengthened by imposing local rotational invariance.  Next, we will move
on to spin systems, and use the above conditions to study entanglement in spin wave states, first
for states containing a small number of spin waves, and then for states at finite temperature.

\section{Examples of states}
One of the advantages of the entanglement conditions in the Introduction is that they allow us to
look at the entanglement between blocks of spins rather than between individual spins.  A standard
approach when studying the entanglement in spin-1/2 systems is to choose two spins and 
calculate their concurrence.  It is, however, quite possible that there is no entanglement between
individual spins, but there is between blocks of spins.  In that case, the method based on concurrence
will fail.  This can be illustrated by an example.  

Let us consider four qubits, i.e. spin-1/2 particles, with qubits $1$ and $2$ in block $a$, and qubits
$3$ and $4$ in block $b$.  Each qubit has an orthonormal basis $\{ |0\rangle , |1\rangle \}$, and
a raising operator $\sigma^{(+)}$ and a lowering operator $\sigma^{(-)}$, where 
$\sigma^{(+)}|0\rangle = |1\rangle$, $\sigma^{(+)}|1\rangle = 0$, and $\sigma^{(-)}=
(\sigma^{(+)})^{\dagger}$.  Let us now consider the four-qubit state
\begin{equation}
|\Psi\rangle = \frac{1}{\sqrt{2}} |00\rangle_{12}|00\rangle_{34} + \frac{1}{2}( |01\rangle_{12}
|10\rangle_{34} + |10\rangle_{12}|01\rangle_{34} ).
\end{equation}
Tracing out qubits $2$ and $4$ we find the reduced density matrix for qubits $1$ and $3$
\begin{equation}
\rho_{13} = \frac{1}{2}|00\rangle_{13}\langle 00| + \frac{1}{4} (|01\rangle_{13}\langle 01|
+ |10\rangle_{13}\langle 10| ) ,
\end{equation}
which is separable.  So, if we just look at qubits $1$ and $3$, i.e. one qubit in each block, we do 
not see any entanglement.  However, setting 
\begin{equation}
 J_{a-} = \sigma_{1}^{(-)} + \sigma_{2}^{(-)}, 
 \hspace{5mm} J_{b-} = \sigma_{3}^{(-)} + \sigma_{4}^{(-)} ,
\end{equation}
we find that 
\begin{equation}
\langle J_{a-}J_{b-}\rangle = \frac{1}{\sqrt{2}},
  \hspace{5mm} \langle J_{a+}J_{a-}\rangle 
 \langle J_{b+}J_{b-}\rangle  = 
\frac{1}{4} ,
\end{equation}
so that the second entanglement condition in Eq.\ (\ref{entcond}) is satisfied.  Therefore, by looking
at entanglement between blocks, we see that the state is, in fact, entangled.

We first will proceed to examine two more complicated types of entangled states in order to see 
whether our entanglement  conditions can show that these states are indeed entangled.  In the 
first case, we will apply the conditions to both individual and collective spins in order to see which
method yields a more sensitive test of entanglement.

\subsection{Correlated sets of qubits}
Suppose we have $2n$ qubits.  We will divide the qubits into two blocks of
$n$ qubits each, and within each block, we will consider only those states of total spin
$j=n/2$.  In particular, we want to examine states of the form
\begin{equation}
\label{states1}
|\Psi\rangle = \sum_{m=-j}^{j} c_{m} |j,m\rangle_{a}\otimes |j,m\rangle_{b}
\end{equation}
with $j=n/2$, and the state with subscript $a$ referring to the first block and the state with subscript
$b$ referring to the second.  This is clearly an entangled state, and we want to see whether the
entanglement conditions we have proposed will detect the entanglement.  We will do this in two
different ways.  First, we will apply the second entanglement condition in Eq.\ (\ref{entcond})
to the collective spin of each block.  Next, we will choose one qubit from each block and apply
the same condition to those two qubits.

The calculations for the condition using the collective spins is straightforward, and we find
\begin{eqnarray}
\langle \Psi |J_{a-}J_{b-}|\Psi \rangle & = & \sum_{m=-j+1}^{j}(j+m)(j-m+1)c^{\ast}_{m-1}c_{m} ,
\nonumber \\
\langle \Psi |J_{a+}J_{a-}|\Psi\rangle & = & \langle\Psi | J_{b+}J_{b-}|\Psi\rangle 
=\sum_{m=-j}^{j} |c_{m}|^{2} (j+m)(j-m+1)  .
\end{eqnarray}
Therefore, the second entanglement condition in Eq.\ (\ref{entcond}) becomes
\begin{equation}
\label{collective}
\left| \sum_{m=-j+1}^{j}(j+m)(j-m+1)c^{\ast}_{m-1}c_{m} \right| > \sum_{m=-j}^{j} |c_{m}|^{2} (j+m)(j-m+1) .
\end{equation}
One possible choice of $c_{m}$ is to set $c_{m}=\eta x^{j+m}$, for some $x>0$ and $\eta$ an 
appropriate normalization constant.  This gives us
\begin{equation}
\left| \sum_{m=-j+1}^{j}(j+m)(j-m+1)x^{2(j+m)-1} \right| > \sum_{m=-j+1}^{j} (j+m)(j-m+1) x^{2(j+m)} .
\end{equation}
This condition is clearly satisfied when $x<1$, but not satisfied for $x>1$.

Now let us see what happens if we just look at one qubit in each block.  Let the qubit in the first block
be qubit $1$ and the one in the second block be qubit $2$, and we will assume that each of these
qubits is the first one in its respective block.  Let us call the spin-down state of an
individual qubit $|0\rangle$ and the spin-up state $|1\rangle$.  The basis states for each block
are $n$-fold tensor products of spin-up and spin-down states for each qubit in the block.
The state $|j,m\rangle$ of $n$
qubits with $j=n/2$ is the symmetric linear combination of all basis states in which there are
$j+m$ ones and $j-m$ zeroes.  There are $\left( \begin{array}{c} 2j \\ j+m \end{array}\right)$ such
states. The operator $\sigma_{1}^{(+)}\sigma_{1}^{(-)}$, where  
$\sigma_{1}^{(+)}$ and $\sigma_{1}^{(-)}$
are the raising and lowering operators for qubit $1$, is just the projection onto states in which
the state of the first qubit is $|1\rangle$.  There are $\left( \begin{array}{c} 2j-1 \\ j+m-1 \end{array}\right)$
states, and this implies that
\begin{equation}
\langle j,m | \sigma_{1}^{(+)}\sigma_{1}^{(-)} |j,m\rangle = \frac{\left( \begin{array}{c} 2j-1 \\ j+m-1 
\end{array}\right)}{\left( \begin{array}{c} 2j \\ j+m \end{array}\right)} =\frac{j+m}{2j} .
\end{equation}  This implies that
\begin{equation}
\langle \Psi | \sigma_{1}^{(+)}\sigma_{1}^{(-)} |\Psi\rangle = \sum_{m=-j}^{j} |c_{m}|^{2} \frac{j+m}{2j} .
\end{equation}
The expression for $\langle \Psi | \sigma_{2}^{(+)}\sigma_{2}^{(-)} |\Psi\rangle$ is identical.

We now want to compute $(\,_{a}\langle j,m-1|\otimes \,_{b}\langle j,m-1 |) \sigma_{1}^{(-)}
\sigma_{2}^{(-)} (|j,m\rangle_{a}\otimes |j,m\rangle_{b})$.  The operator $ \sigma_{1}^{(-)} 
\sigma_{2}^{(-)}$ will pick  out the basis states with ones in the first slot of each block.  By reasoning
similar to that above, we have that
\begin{eqnarray}
(\,_{a}\langle j,m-1|\otimes \,_{b}\langle j,m-1 |) \sigma_{1}^{(-)} \sigma_{2}^{(-)} (|j,m\rangle_{a}
\otimes |j,m\rangle_{b})  \nonumber \\
=  \frac{\left( \begin{array}{c} 2j-1 \\ j+m-1 \end{array}\right)^{2}}
{\left( \begin{array}{c} 2j \\ j+m-1 \end{array}\right) \left( \begin{array}{c} 2j \\ j+m \end{array}\right)}
\nonumber  \\
 =  \frac{(j+m)(j-m+1)}{(2j)^{2}} .
\end{eqnarray} 
This gives us that
\begin{equation}
\langle\Psi |  \sigma_{1}^{(-)} \sigma_{2}^{(-)} |\Psi\rangle = \sum_{m=-j+1}^{j} c_{m-1}^{\ast}
c_{m} \frac{(j+m)(j-m+1)}{(2j)^{2}} .
\end{equation}
Finally, the second entanglement condition in Eq.\ (\ref{entcond}) with $A=\sigma_{1}^{(-)}$ and
$B=\sigma_{2}^{(+)}$ becomes
\begin{equation}
\label{singlespin}
\left| \sum_{m=-j+1}^{j}(j+m)(j-m+1)c^{\ast}_{m-1}c_{m} \right| > (2j) \sum_{m=-j}^{j} |c_{m}|^{2} (j+m) .
\end{equation}
Comparing  Eqs.\ (\ref{collective}) and (\ref{singlespin}) we note that $2j \geq j-m+1$ if $m\geq -j+1$,
which is the entire range of the sum.  This implies that at least for states of the type in Eq.\ (\ref{states1}),
the collective spin condition is stronger, that is, it will be satisfied by more states, than the condition for individual spins.

\subsection{Angular momentum intelligent states}
We first want to find some spin states that satisfy our entanglement conditions.  One possibility is to
find states in which the spins of the two subsystems are highly correlated, and states in which the
uncertainty of the sum (or difference) of the two spins is small will satisfy this condition. 

Let us begin by looking at the uncertainty relation for the total spin,
\begin{eqnarray}
    \Delta (J_{a1}+J_{b1})  \Delta (J_{a2}+J_{b2})\geq
    \frac {1}{2}|(J_{a3}+J_{b3})| ,
\end{eqnarray}
where $J_{a1}$, $J_{a2}$, and $J_{a3}$ are the components of $\mathbf{J}_{a}$, and 
 $J_{b1}$, $J_{b2}$, and $J_{b3}$ are the components of $\mathbf{J}_{b}$.  We would like to find 
the states which satisfy this relation as an equality. These states were first found in \cite{aragone},
and here we will follow the treatment given in \cite{hillery4}.  These  satisfy the eigenvalue equation
\begin{eqnarray}
\label{intell-eigen}
    [(J_{1a}+J_{1b})+i\lambda(J_{2a}+J_{2b})]|\Psi \rangle = \beta
    |\Psi \rangle  ,
\end{eqnarray}
where $\lambda$ is real.  This equation implies that
\begin{equation}
\langle \Psi | (J_{a1}+J_{b1})|\Psi\rangle = {\rm Re}(\beta ), \hspace{5mm}
\langle \Psi | (J_{a2}+J_{b2})|\Psi\rangle = (1/\lambda ){\rm Im}(\beta ),
\end{equation}
and
\begin{equation}
[ \Delta (J_{a1}+J_{b1}) ]^{2}= \frac{\lambda}{2} \langle J_{a3}+J_{b3}\rangle \hspace{5mm}
[ \Delta (J_{a2}+J_{b2}) ]^{2}= \frac{1}{2\lambda} \langle J_{a3}+J_{b3}\rangle  .
\end{equation}
From these equations, we see that when $\lambda$ is small, $J_{a1}$ and $-J_{b1}$ are highly 
correlated, and when it is large, $J_{a2}$ and $-J_{b2}$ are highly correlated.  
These states are spin analogs of two-mode squeezed state for light.

In order to solve Eq.\ (\ref{intell-eigen}), we first define a state 
\begin{equation}
|\Psi^{\prime}\rangle =e^{i\theta (J_{a1}+J_{b1})}|\Psi \rangle ,
\end{equation}
and insert the resulting expression for $|\Psi\rangle$ into Eq.\ (\ref{intell-eigen}) to give an
equation for 
$|\Psi^{\prime}\rangle$ 
\begin{eqnarray}
    \{(J_{a1}+J_{b1})+i \lambda [(J_{a2}+J_{b2}) \cos{\theta}-(J_{a3}+J_{b3})
    \sin{\theta}]\}|\Psi^{\prime}\rangle = \beta |\Psi^{\prime} \rangle .
\end{eqnarray}
Now choose $\lambda=- 1/\cos {\theta}$, and $\theta$ to be in
the range $\pi\geq \theta \geq \pi/2 $, which implies that $\lambda >1$, and
\begin{equation}
[ (J_{a-}+J_{b-})-i\sqrt{\lambda^{2}-1}(J_{a3}+J_{b3})]|\Psi^{\prime}\rangle = \beta |\Psi^{\prime}\rangle .
\end{equation}
We now expand $|\Psi^{\prime}\rangle$ as 
\begin{equation}
|\Psi^{\prime} \rangle= \sum_{n,m=-j}^{j} C_{nm}|n,m\rangle ,
\end{equation}
where we have set $|n,m\rangle = |j,n\rangle \otimes |j,m\rangle$.  If we assume, for simplicity, that
$C_{n,m}=0$, unless $n=m$, our equation for $|\Psi^{\prime}\rangle$ reduces to the
recurrence relation
\begin{equation}
C_{m+1,m+1}=\frac{ \beta +2mi \sqrt{\lambda^{2}-1} }{(j+m+1)(j-m)} C_{m,m}, \hspace{5mm} m < j,  
\end{equation}
\begin{equation}
[\beta +2ij \sqrt {\lambda^2-1}]C_{j,j}=0, \hspace{5mm} m=j  .
\end{equation}

From the second equation, we see that either $\beta =-2ij \sqrt
{\lambda^2-1}$, or $C_{j,j}=0$.  If $C_{j,j}=0$, then it must be the case that $\beta =-2m_0 i \sqrt
{\lambda^2-1}$ for some $m_0$. So, 
\begin{equation}
C_{m,m}= (-2i \sqrt{\lambda^2-1})^{j+m} \frac{(m_0+j)!(j-m)!}{(m_0-m)!(j+m)! (2j)!}C_{-j,-j}  , 
\end{equation}    
for $m\leq m_{0}$ and $C_{mm}=0$ for $m>m_{0}$.  After grouping the m-independent constants 
$\frac{(m_0+j)!}{(2j)!}C_{-j,-j}$ into $C_{j,m_0}(\lambda)$, we have
\begin{eqnarray}
|\Psi (j,m_0,\lambda) \rangle & = & C_{j,m_0}(\lambda)e^{-i\theta(J_{1a}+J_{1b})}  \nonumber  \\
& &  \sum_{m=-j}^{m_0}(-2j \sqrt {\lambda^2-1})^{j+m} \frac{(j-m)!}{(m_0-m)!(j+m)!}|m,m \rangle  .
\end{eqnarray}
We want to see if there is some range of parameters for which  $ |\langle J_{a-}J_{b-} \rangle|^2 > \langle J_{a+}J_{a-} \rangle  \langle J_{b+}J_{b-} \rangle$, but for these states 
$ \langle J_{a+}J_{a-} \rangle = \langle J_{b+}J_{b-} \rangle $, so we just need to show that 
$|\langle J_{a-}J_{b-} \rangle| >\langle J_{a+}J_{a-} \rangle$.  
We find that
\begin{eqnarray}
\langle J_{a+}J_{a-} \rangle & = &  |C_{j,m_0}(\lambda)|^2 \sum_{m=-j}^{m_0}
[4(\lambda^2-1)]^{j+m} [\frac{(j-m)!}{(m_0-m)!(j+m)!}]^2  \nonumber \\
 & & \times   
 \left\{  \frac {\lambda^2+1}{2\lambda ^2} [ j(j+1)-m^2]- \frac {m}{\lambda}  \right\} ,
\end{eqnarray}
and
\begin{eqnarray}
\langle J_{a-}J_{b-} \rangle & = & |C_{j,m_0}(\lambda)|^2 \sum_{m=-j}^{m_0}
[4(\lambda^2-1)]^{j+m} [\frac{(j-m)!}{(m_0-m)!(j+m)!}]^2 \nonumber  \\
& & \times 
   \left[ \frac {2i \sqrt {\lambda^2+1}}{\lambda} (m_0-m)- \frac {(\lambda^2-1)m^2}{\lambda^2} \right] .
\end{eqnarray}
Consider the simple case in which  $j=1$,$m_0=-1$,and $m=-1$, so that the sum has only one
term.  The entanglement condition becomes
\begin{eqnarray}
    \left|\frac{\lambda^2-1}{\lambda^2} \right|
    >\frac{(\lambda+1)^2}{2\lambda^2}  ,
\end{eqnarray}
and the state is entangled if 
$\lambda >3$.

\section{Local rotational invariance}
Entanglement is not affected by local unitary transformations, and so, ideally, we would like our entanglement conditions to be invariant under local unitaries as well.   It is not always possible to
accomplish this, but we can sometimes obtain invariance under a subgroup of the group of
local unitary transformations.  In Ref.\ \cite{hillery3} in which entanglement between field modes
was considered, it was possible to find entanglement conditions that are invariant under local
Gaussian transformations of the field modes.  These new conditions were stronger than the original
ones, that is they detect entanglement in a larger set of states.  Thus, making the conditions
invariant under a subset of local unitary transformations strengthens them. 

For the entanglement conditions we are considering in this paper, the obvious group of local 
unitaries consists of local rotations.  Under the action of the rotation $R(\alpha,\beta,\gamma)
=e^{-i\alpha J_1}e^{-i\beta J_2}e^{-i\gamma J_3}$, we have that
\begin{eqnarray}
 R^{-1}J_+R 
& = & [\frac{1}{2}(\cos{\alpha}+\cos{\beta})+\frac{i}{2}\sin{\alpha}\sin{\beta}]
 e^{i\gamma}J_{+} \nonumber \\
 & &  +[\frac{1}{2}(\cos{\beta}-\cos{\alpha})+\frac{i}{2}\sin{\alpha}\sin{\beta}]e^{-i\gamma}J_{-}
\nonumber \\
& & +[1-i\sin{\alpha}\sin{\beta}]e^{i\gamma}J_3 .
\end{eqnarray}
Now suppose we start with the entanglement condition Eq.\ (\ref{entcongen1}), and we want to
find from it a condition that is invariant under local rotations of the $a$ system (finding a condition
that is invariant under local rotations 
of both $a$ and $b$ subsystems
is possible, but it results in a $9\times 9$ matrix, which is
rather unwieldy).  We note that what the local rotation on subsystem $a$ does is to send both
$J_{a+}$ and $J_{a-}$ into linear combinations of $J_{a+}$, $J_{a-}$, and $J_{a3}$.  This
suggests that we set  $A=c_1^*J_{a-}+c_2^*J_{a+}+c_3^* J_{a3}$ and $B=J_{b-}$ in 
Eq.\ (\ref{entcongen1}).  The entanglement condition can then be written in the form
\begin{equation}
\label{rotation}
 \left( \begin{array}{ccc}
 c_{1}^{\ast}  & c_{2}^{\ast} &  c_{3}^{\ast}  \end{array} \right) M
 \left( \begin{array}{c} c_{1} \\ c_2 \\ c_3 \end{array} \right)  > 0  ,
\end{equation}
where $M$ is a $3\times 3$ matrix, whose elements are linear combinations of expectation values
of products of angular momentum operators.  In particular, 
\begin{eqnarray}
 && M_{11} = |\langle J_{a-}J_{b+}\rangle |^{2} 
 - \langle J_{a+}J_{a-}J_{b+}J_{b-}\rangle, 
\nonumber \\
 && M_{12} = \langle J_{a-}J_{b+}\rangle^{\ast}\langle J_{a+}J_{b+}\rangle 
 - \langle J_{a+}^{2}J_{b+}J_{b-}\rangle ,
\nonumber \\
 && M_{13} = \langle J_{a-}J_{b+}\rangle^{\ast}\langle J_{a3}J_{b+}\rangle 
 - \langle J_{a+}J_{a3}J_{b+}J_{b-}\rangle ,
\nonumber \\
 && M_{21}= \langle J_{a+}J_{b+}\rangle^{\ast}
      \langle J_{a-}J_{b+}\rangle 
  - \langle J_{a-}^{2}J_{b+}J_{b-}\rangle,  
\nonumber \\
 && M_{22}=|\langle J_{a+}J_{b+}\rangle |^{2} 
   - \langle J_{a-}J_{a+}J_{b+}J_{b-}\rangle ,
\nonumber \\
 && M_{23} = \langle J_{a+}J_{b+}\rangle^{\ast}\langle J_{a3}J_{b+}\rangle 
 -\langle J_{a-}J_{a3}J_{b+}J_{b-}\rangle  ,
\nonumber \\
 && M_{31} = \langle J_{a3}J_{b+}\rangle^{\ast}\langle J_{a-}J_{b+}\rangle 
 - \langle J_{a3}J_{a-} J_{b+}J_{b-}\rangle , 
\nonumber \\
 && M_{32}= \langle J_{a3}J_{b+}\rangle^{\ast} \langle J_{a+}J_{b+}\rangle
 -\langle J_{a3}J_{a+}J_{b+}J_{b-}\rangle ,
\nonumber \\
 && M_{33}= |\langle J_{a3}J_{b+}\rangle |^{2} 
 - \langle J_{a3}^{2}J_{b+}J_{b-}\rangle .
%
\end{eqnarray}
If we change the state by a local rotation of system $a$,
the effect on Eq.\ (\ref{rotation}) is only to change the values of $c_{1}$, $c_{2}$ and $c_{3}$.  This
follows from the fact that when $A$ is conjugated by the rotation $R_a$, the form of the operator
stays the same, that is, it is a linear combination of $J_{a+}$, $J_{a-}$, and $J_{a3}$, but the 
coefficients multiplying the operators change.  If the matrix $M$ has a positive eigenvalue, then
we can find values of  $c_{1}$, $c_{2}$ and $c_{3}$ so that the above condition is satisfied, simply
by choosing them to be the components of the vector corresponding to the positive eigenvalue.  
Therefore, our new entanglement condition becomes that $M$ has a positive eigenvalue, and this
condition is invariant under local rotations on system $a$.

Let us show that this new condition is stronger than our original condition.  If the state we are
considering is
\begin{eqnarray}
    | \Psi \rangle = \frac {1}{\sqrt{2}}(|-j,-j+1\rangle + |-j+1, -j \rangle ) ,
\end {eqnarray}
then
\begin{eqnarray}
    M=
    \left(    \begin{array}{ccc}
    j^2 & 0 & 0 \\
    0 & -2j^2 & 0 \\
    0 & 0 & -j^3
    \end{array}
    \right)  .
\end{eqnarray}
Noting that $j^2$ is positive, we see that the state is entangled.  Because this condition is
invariant under rotations of system $a$, it would also show that the state $R_{a}\otimes I_{b}
|\Psi\rangle$ is entangled.

Now, let us see what happens if we apply our original condition to the state $R_{a}\otimes I_{b}
|\Psi\rangle$.  We begin by finding
\begin{eqnarray}
    \Bigl | \langle \Psi
    |R_a^{-1}J_{a+}R_aJ_{b-}|\Psi\rangle\Bigr |^2
    =\frac{j^2}{4}[(\cos{\alpha}+\cos{\beta})^2+\sin^2{\alpha}\sin^2{\beta}] ,
\end{eqnarray}
and
\begin{eqnarray}
 \langle \Psi|R_a^{-1}J_{a+}J_{a-}R_a J_{b+}J_{b-}|\Psi\rangle
& = & \frac{j^2}{2}[(\cos{\beta}-\cos{\alpha})^2+\sin^2{\alpha}\sin^2{\beta}]  \nonumber \\
& & + j^3(1+\sin^2{\alpha}\sin^2{\beta}).
\end{eqnarray}
Therefore, the state is entangled according to the old condition if
\begin{equation}
 \cos{\alpha}\cos{\beta}
    > j(1+\sin^2{\alpha}\sin^2{\beta}) + \frac{1}{4}
      (\cos{\alpha}\cos{\beta}-1)^2.
\end{equation}
This condition can be satisfied for only a limited range of $\alpha$ and $\beta$ if $j$ is small, and 
it cannot be satisfied at all if 
$j\ge 1$, which actually allows only for $j=1/2$.
Therefore, our new condition, which is invariant under rotations
of system $a$, is considerably more powerful in that it detects entanglement in a much larger set of
states.

\section{Spin waves}
The low-lying energy states of a system of spins coupled by exchange interactions are wavelike, as shown originally by Bloch for ferromagnets. The waves are called spin waves, and they correspond
to excitations of definite energy called magnons.   We will study the entanglement between spins,
and blocks of spins for magnon states in a ferromagnet.  We will first examine entanglement in states
containing a small number of magnons, and then go on to study the case of a ferromagnet at low, but
finite, temperature. 

\subsection{Small number of magnons}
The Hamiltonian describing spins on a lattice interacting via a nearest-neighbor exchange interaction
and an externally applied magnetic field is \cite{kittel,feynman}
\begin{equation}
H=-J\sum_{\mathbf{j},\mathbf{\delta}}\mathbf{S}_{\mathbf{j}}\cdot 
\mathbf{S}_{\mathbf{j}+\mathbf{\delta}}-2\mu_{0}H_{0}\sum_{\mathbf{j}} S_{\mathbf{j}z} ,
\end{equation}
where the vectors $\mathbf{\delta}$ connect the spin at site $\mathbf{j}$ with its nearest neighbors on 
a bravais lattice, $J$ is the exchange integral, which is assumed to be positive, $\mu_0=(g/2)\mu_B$ 
is the magnetic moment of the atoms, and $\mathbf{S}_{\mathbf{j}}$ is the spin angular momentum
operator of the atom at $\mathbf{j}$.  $H_0$ is the intensity of a static magnetic field directed along the
$z$ axis, and we will take the limit as $H_0\rightarrow 0^{+}$ to make the magnetic moments line up
along the positive $z$ axis when the system is in the ground state
$|\Omega\rangle$.
The $z$ component of the total spin, $\mathcal{S}_z=\sum_j S_{jz}$ is a constant of the motion, and the ground state of the system simply
has all of the spins pointing in the $+z$ direction.  

For the case of a small number of spin waves, let us consider a line of $N$ spins with periodic
boundary conditions (the spin at $N+1$ is the same as the spin at $1$).  If the atoms have a spin
of $1/2$, then the state containing a single magnon is a linear combination of states with one
spin flipped
\begin{equation}
|\Psi\rangle = \frac{1}{\sqrt{N}}\sum_{j=1}^{N}e^{ikja}\sigma_j^{(-)}|\Omega\rangle ,
\end{equation}
where $a$ is the spacing between spins, and $k=2\pi n/(Na)$, where $n$ is an integer in the range
$-(N/2)< n \leq (N/2)$.  
%
The operator $\sigma_j^{(-)}$ is the spin lowering operator for the spin at
site $j$, that is it maps the spin up state at site $j$ to the spin down state at the same site.  

Now let us examine the entanglement of this state using  Eq.\ (\ref{entcongen1}).  Let
\begin{eqnarray}
A & = & S_{1+} = \sum_{j=1}^{m} \sigma_j^{(+)} ,
\nonumber \\
B & = & S_{2+} = \sum_{j=L+1}^{L+m}\sigma_j^{(+)}, 
\end{eqnarray}
where $m$ is a number such that $2m<N$.  This will allow us to see if there is entanglement
between two 
blocks of spins each of size $m$
and distanced from each other by $(L-m)$ spins.
Our state is entangled if
\begin{equation}
|\langle S_{1+}S_{2-}\rangle |^{2} > \langle S_{1-}S_{1+}S_{2-}S_{2+}\rangle .
\end{equation}
For the single magnon state above, the right-hand side is zero, so as long as the left-hand side is
non-zero, we can say that the blocks of spins are entangled.  In fact, we find that
\begin{equation}
\langle S_{1+}S_{2-}\rangle =\frac{1}{N} 
\sum_{j_{1}=1}^{m} \sum_{j_{2}=L+1}^{L+m} e^{ika(j_{2}-j_{1})} .
\end{equation}
If the size of the blocks is small compared to the wavelength of the spin wave, the term in the sum will 
all have approximately the same phase, and will add coherently.  This would show that the blocks of
spins are entangled for this state.

If we want to look at more than one magnon, more sophisticated techniques are required.  We will
make use of the Holstein-Primakoff transformation, which expresses the spin operators in terms
of boson creation and annihilation operators, and allows us to approximately diagonalize the
Hamiltonian.  The Holstein-Primakoff transformation of the spin operator $\mathbf{S}_{\mathbf{j}}$
to boson creation and annihilation operators $a_{\mathbf{j}}^{\dagger}$, $a_{\mathbf{j}}$ is given by
\begin{eqnarray}
S_{\mathbf{j}+} & = & S_{\mathbf{j}x}+iS_{\mathbf{j}y}=(2S-a_{\mathbf{j}}^{\dagger}
a_{\mathbf{j}})^{1/2}a_{\mathbf{j}} , \nonumber  \\
S_{\mathbf{j}-} & = & S_{\mathbf{j}x}-iS_{\mathbf{j}y}= a_{\mathbf{j}}^{\dagger}(2S
-a_{\mathbf{j}}^{\dagger}a_{\mathbf{j}})^{1/2} , \nonumber \\
S_{\mathbf{j}z} & = & S- a_{\mathbf{j}}^{\dagger}a_{\mathbf{j}} ,
\end{eqnarray}
where
\begin{equation}
[a_{\mathbf{j}}, a_{\mathbf{l}}^{\dagger} ]=\delta_{\mathbf{j},\mathbf{l}} .
\end{equation}
If we consider only situations in which the number of flipped spins is small compared to the total
number of spins, we can expand the square roots and keep only the first terms in the expansion.
In addition we make a transformation from the spin operators, 
 $a_{\mathbf{j}}^{\dagger}$ and 
$a_{\mathbf{j}}$, to the magnon variables, $b_{\mathbf{k}}^{\dagger}$and $b_{\mathbf{k}}$, defined by
\begin{equation}
b_{\mathbf{k}}=N^{-1/2}\sum_{\mathbf{j}} e^{i\mathbf{k}\cdot \mathbf{r}_{\mathbf{j}}} a_{\mathbf{j}} ,
\end{equation}
where $\mathbf{r}_{\mathbf{j}}$ is the position of spin $\mathbf{j}$.  The magnon operators satisfy boson
commutation relation:
\begin{equation}
[b_{\mathbf{k}}, b_{\mathbf{k}^{\prime}}^{\dagger} ] = \delta_{\mathbf{k},\mathbf{k}^{\prime}},
\hspace{5mm} [b_{\mathbf{k}}, b_{\mathbf{k}^{\prime}} ] =0 .
\end{equation}
When the number of flipped spins is much less than $N$, the Hamiltonian is diagonal in the magnon
operators,
\begin{equation}
H=\sum_{\mathbf{k}}\omega_{\mathbf{k}} b_{\mathbf{k}}^{\dagger} b_{\mathbf{k}} ,
\end{equation}
where 
\begin{equation}
\omega_{\mathbf{k}} = 2JzS(1-\gamma_{\mathbf{k}}) + 2\mu_{0}H_{0} ,
\end{equation}
and
\begin{equation}
\gamma_{\mathbf{k}} = \frac{1}{z}\sum_{\mathbf{\delta}}e^{i\mathbf{k}\cdot \mathbf{\delta}} .
\end{equation}
As was mentioned before, we will work in the limit $H_{0}\rightarrow 0^{+}$, so that in the ground
state the spins are lined up along the $z$ axis.  
In these equations a center of symmetry is assumed so that $\gamma_{\mathbf{k}}=
\gamma_{-\mathbf{k}}$, and $z$ is the number of nearest neighbors each spin has.

Now we are in a position to consider the two-magnon state.  We shall again consider the 
one-dimensional case, that is $N$ spins in a line.  We want to study the entanglement of the state 
\begin{equation}
|\Psi \rangle =b_{k_1}^\dagger
b_{k_2}^\dagger|0\rangle=\frac{1}{N}\sum_{u,v}e^{-ik_1x_u}e^{-ik_2x_v} 
a_u^\dagger a_v^\dagger|0 \rangle 
\end{equation}
for $k_1\neq k_2$.
We shall examine the entanglement between two blocks consisting of $m$ spins each, one
beginning at spin $1$ and the other beginning at spin $L$, so that the blocks are separated
by $L-m$ spins.  Therefore, we choose 
\begin{eqnarray}
\label{AB_def}
   A=\sqrt{2S}\sum_{j=1}^m \ a_{j},\qquad
   B=\sqrt{2S} \sum_{j=L+1}^{L+m} \ a_{j},
\end{eqnarray}
in Eq.\ (\ref{entcongen1}). We find
\begin{equation}
\langle A^\dagger A B^\dagger B \rangle=\frac{4S^2}{N^2}\{2xy+2xy\cos{[La(k_1-k_2)]}\} ,
\end{equation}
where 
$x= [\cos (k_{1}ma)+1]/[\cos k_{1}a+1]$, and 
$y= [\cos (k_{2}ma)+1]/[\cos k_{2}a+1]$, and 
\begin{equation}
|\langle A^\dagger B \rangle|^2=\frac{4S^2}{N^2}
   \{x^2+2xy\cos{[La(k_1-k_2)]}+y^2\}  .
\end{equation}
Therefore,  the state is entangled if
\begin{equation}
\label{2mag}
  (x-y)^2>0,
\end{equation}
%
which is true as long as $x\neq y$. 
One situation where $x=y=1$ is when the block size is one $m=1$, implying 
that the condition (\ref{entcongen1}) does not detect entanglement between 
individual spins in the two-magnon state. Recall that 
$k_1 a = \pi 2n_1/N$ and $k_2 a = \pi 2n_2/N$. If the block size $m$
is such that $m 2n_1/N=2l_1+1$ and $m 2n_2/N=2l_2+1$ where $l_1$ and 
$l_2$ are integers, hence $x=y=0$ and no entanglement is found according
to the inequality (\ref{2mag}).
The condition (\ref{2mag}) indicates that in the ideal zero-temperature 
two-magnon state, entanglement is found regardless of how far the two blocks are separated. This no longer occurs in the more realistic non-zero 
temperature state we are going to investigate below.

\subsection{Finite temperature}
Now that we have seen that the entanglement condition, Eq.\ (\ref{entcongen1}), is useful in detecting
entanglement in states consisting of a few magnons, let us see whether it can also detect entanglement
in a system of ferromagnetically interacting spins at a finite temperature, $T$.  The density matrix for 
the system is now given by
\begin{equation}
 \rho = \frac{1}{Z} e^{-\beta H} ,
\end{equation}
where $ \beta=1/(k_{B} T)$ and $k_{B}$ is Boltzmann's constant.  The partition function of the 
system, $Z$ is given by
\begin{equation}
Z = {\rm Tr} (e^{-\beta H})= \prod_{\bf k}  \sum_{n_{\bf k}}
           e^{-\beta\omega_k n_{\bf k}}
         = \prod_{ {\bf k}} \frac{1}{(1-e^{-\beta\omega_k})} ,
\end{equation}
and $n_{\mathbf{k}}$ is the number of magnons with wave vector $\mathbf{k}$.

We first look for entanglement between two individual spins having
radius vectors ${\bf r}_1$ and ${\bf r}_2$ by employing 
the inequality in Eq.\ (\ref{entcongen1}) with
\begin{eqnarray}
\label{f10}
      A = S_{j_{1}+}=\sqrt{2S}a_1, \qquad B = S_{j_{2}+}=\sqrt{2S}a_2.
\end{eqnarray}
We then have that 
\begin{eqnarray}
\label{f11}
    &&  \langle AB^\dagger \rangle 
     = \langle S_{j_1+} S_{j_2-} \rangle
\nonumber\\
    && = \frac{2S}{N} \sum_{{\bf k}_1}\sum_{{\bf k}_2} 
            e^{-i{\bf k}_1 \cdot {\bf r}_1+i{\bf k}_2 \cdot {\bf r}_2}
               {\rm Tr}(b_{{\bf k}_1} b_{{\bf k}_2}^\dagger \rho) .
\end{eqnarray}
Using the relationship $\sum_{n=0}^{\infty} (n+1)x^n = 1/(1-x)^2$, one obtains
\begin{eqnarray}
\label{f13}
      {\rm Tr}(b_{{\bf k}_1} b_{{\bf k}_2}^\dagger \rho) = 
         \frac{\delta_{{\bf k}_1,{\bf k}_2} }{1-e^{-\beta\omega_{k_1}}},
\end{eqnarray}
so that 
\begin{eqnarray}
\label{f14}
      \langle AB^\dagger \rangle 
     = \frac{2S}{N} \sum_{{\bf k}_1} 
       \frac{e^{-i{\bf k}_1 \cdot
         ({\bf r}_1-{\bf r}_2)}}{1-e^{-\beta\omega_{k_1}}}.
\end{eqnarray}
This gives us the left-hand side of our inequality, and we now need to find the right-hand side.  Using
\begin{eqnarray}
\label{f15}
     && \langle n_{{\bf k}_1},n_{{\bf k}_2},
           n_{{\bf k}_3},n_{{\bf k}_4},\ldots|
        b_{{\bf k}_1}^\dagger b_{{\bf k}_2} 
          b_{{\bf k}_3}^\dagger b_{{\bf k}_4} 
         |n_{{\bf k}_1},n_{{\bf k}_2},
          n_{{\bf k}_3},n_{{\bf k}_4},\ldots \rangle =
\nonumber\\
     &&\delta_{{\bf k}_1,{\bf k}_2}  
       \delta_{{\bf k}_3,{\bf k}_4} n_{{\bf k}_1}n_{{\bf k}_3} 
    + \delta_{{\bf k}_1,{\bf k}_4}  
       \delta_{{\bf k}_2,{\bf k}_3} n_{{\bf k}_1}(n_{{\bf k}_2}+1),
\end{eqnarray}
we obtain for the right hand side of Eq. (\ref{entcongen1})
\begin{eqnarray}
\label{f16}
    &&  \langle A^\dagger A B^\dagger B \rangle 
     = \langle S_{j_{1}-} S_{j_{1}+} S_{j_{2}-} S_{j_{2}+} \rangle
\nonumber\\
    && \hspace{-1cm} = \left( \frac{2S}{N} \right)^2 
        \sum_{{\bf k}_1,{\bf k}_2}
      \biggl[
       \frac{e^{-\beta(\omega_{k_1}+\omega_{k_2})}}
          {(1-e^{-\beta\omega_{k_1}})(1-e^{-\beta\omega_{k_2}})}
       + \frac{e^{i({\bf k}_1-{\bf k}_2) \cdot ({\bf r}_1-{\bf r}_2)}
                  e^{-\beta\omega_{k_1}} }
          {(1-e^{-\beta\omega_{k_1}})(1-e^{-\beta\omega_{k_2}})}
         \biggr].
\end{eqnarray}
Equation (\ref{f16}) shows that $\langle A^\dagger A B^\dagger B \rangle$ can be
separated into two parts: the first one represents the self
correlation of the particles and is distance independent, while the second
represents interparticle correlations and depends on the 
distance between the particles.

Let us now specialize to a cubic lattice with lattice constant $a$ and 
$z=6$.   If $|{\bf k} a\, | \ll 1$, then 
\begin{eqnarray}
\label{f17}
    1-\gamma_{{\bf k}} = 1 - \frac{1}{3}(\cos k_x a+\cos k_y a+\cos k_z a)
     \simeq \frac{1}{6} k^2 a^2  , 
\end{eqnarray}
and the magnon energy can be expressed as $ \omega_k = D k^2$, where $D = 2JS a^2$.
To tackle the sums over ${\bf k}_j$ we note that
\begin{eqnarray}
\label{f19}
         - \frac{\pi}{a}  < k_j \le \frac{\pi}{a} 
\end{eqnarray}
and approximate the cube by a sphere, so that $k_j \le \sqrt{3} \pi/a$.
We replace the sums by integrals in a spherical coordinate system,
and our entanglement inequality, Eq.\  (\ref{entcongen1}) becomes, upon using Eqs. (\ref{f14}) and
(\ref{f16}) and carrying out the angular integrations,
%
\begin{equation}
\label{f20}
     Q = I_1^2 - (I_2^2 + I_1 I_3)>0,
\end{equation}
where $|\langle AB^\dagger \rangle|^2=I_1^2$, 
$\langle A^\dagger A B^\dagger B \rangle = I_2^2 + I_1 I_3$,
\begin{eqnarray}
\label{f20.1}
     &&I_1 = \int_0^{y_0} dy 
     f\left(\frac{y}{y_0} \frac{\Delta r}{a} \pi\sqrt{3}\right)
     \frac{y^2}{1-e^{-y^2}} ,
\\
\label{f20.2}
     &&I_2 = \int_0^{y_0} dy 
     \frac{y^2e^{-y^2}}{1-e^{-y^2}} ,
\\
\label{f20.3}
     &&I_3 = \int_0^{y_0} dy 
     f\left(\frac{y}{y_0} \frac{\Delta r}{a} \pi\sqrt{3}\right)
     \frac{ y^2 e^{-y^2}}{1-e^{-y^2}} ,
\end{eqnarray}
and $f(x)$ is the familiar function
\begin{eqnarray}
\label{f21}
      f(x) = \frac{\sin x}{x}.
\end{eqnarray}
Here $\Delta r = |{\bf r}_1-{\bf r}_2|$ is the interatomic distance,
\mbox{$y_0= \sqrt{\beta D} \sqrt{3}\pi/a$} and the dimensionless integration
variable $y$ is related to the wave vector component $k$ by  
$y = \sqrt{\beta D} k$. 
Due to the presence of the exponentially decaying
factor $e^{-y^2}$ in the numerators of the integrands in $I_2$ and $I_3$, 
small values of $y$, $y\stackrel{<}{\sim} 1$, contribute most to these 
integrals. In the case of $I_3$, the fact that
$y^2 e^{-y^2}/(1-e^{-y^2})$ is a decreasing function, causes that integral
to be positive.

As $T$ increases, the upper limit of the integrals, $y_0$, which is 
proportional to $\frac{1}{\sqrt{T}}$, tends to zero,  and, 
as a result, $e^{-y^2} \rightarrow 1$ and $I_3 \rightarrow I_1$. Hence
the inequality (\ref{f20}) becomes $I_1^2 - (I_2^2 + I_1^2)>0$, which cannot
be fulfilled. This means
that our condition does not show the
existence of entanglement in the high temperature limit, which is consistent with
what we expect on physical grounds, i.e.\ that there is no entanglement at high temperature.  
As the temperature decreases, the upper integral limit $y_0$ increases. 
The integrand in $I_1$ is an oscillating function of $y$ with 
a varying sign and an increasing magnitude, and the sign and value of $I_1$
are determined mostly by the contribution near $y_0$.  For short distances and 
low temperatures, the absolute value
of $I_2$ is typically much larger than those of $I_1$ and $I_3$,
which makes it the leading factor in deciding the sign of $Q$.

\begin{figure}
\includegraphics{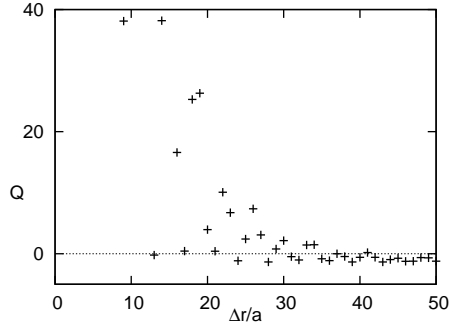}
\caption{
 The quantity $Q$ as a function of the interatomic distance, scaled by the 
lattice constant. A positive $Q$ indicates entanglement. Some large
(positive) values of $Q$ are beyond the scope of the figure.
The temperature is fixed at $\sqrt{2JS/(k_{\rm B}T)}=7$.
}
\label{dx}
\end{figure}

In Fig. \ref{dx} we give a representative example of the distance 
dependence of $Q$. Positive values of $Q$ indicate entanglement. 
It can be seen from Eq. (\ref{f20.1}) that $I_1$ is an oscillating function
of the interatomic distance $\Delta r$, with a damping envelop. This
shows up in the behavior of $Q$: If we allow for continuous values of
$\Delta r/a$ we will see the damped oscillations more clearly. For 
short interatomic distances, entanglement is clearly observed.
$Q$ turns negative for the first time at $\Delta r/a=13$. However, it can
again become positive, meaning the reappearance of detectable entanglement at 
much larger distances before becoming permanently negative.
In Fig. \ref{dx} the temperature is fixed. For lower 
temperatures, the shortest distance at which $Q$ is found to be 
negative and the overall range over which $Q$ is found to be positive  increases.

\begin{figure}
\includegraphics{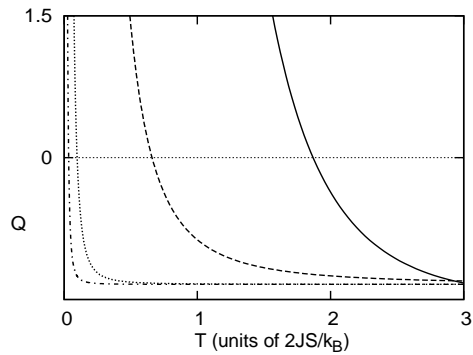}
\caption{
 The quantity $Q$ as a function of the temperature for 
different inter-particle distances $\Delta r/a=$ (a) 1 (solid line), 
(b) 3 (dashed line), (c) 10 (dotted line), (d) 20 (dot-dashed line).
}
\label{temp}
\end{figure}

The temperature dependence of $Q$ is illustrated in Fig. \ref{temp} for
different values of the inter-particle distance. It can be seen that as the temperature increases,  
$Q$ monotonically decreases, and the shorter the interparticle distance,
the later $Q$ crosses into the negative range. In other words, as we would expect on physical
grounds, lower temperatures and shorter inter-particle distances are more favorable for 
entanglement generation, and this parameter region is where our condition shows the 
presence of entanglement.
If we assume some typical parameters for ferromagnets \cite{kittel}
$D\sim 0.5\times 10^{-28}\ {\rm erg\ cm}^2$
and $a\sim 4 \AA$, using $k_{\rm B}=1.38\times 10^{-16}\ {\rm erg\ K}^{-1}$,
the temperatures at which $Q$ turns negative, which is where entanglement is
no longer detected, are 420K, 150K, 20K, and 8K for $\Delta r/a=$ 1 (solid line),
3 (dashed line), 10 (dotted line), and 20 (dot-dashed line), respectively.
Thus when the atoms are closer located, entanglement can be detected at 
higher temperatures. We are, of course, assuming that these temperatures 
are still considerably below the critical temperature, so that the spin-wave 
description remains valid.

Let us now proceed to use the condition (\ref{entcongen1}) to investigate 
entanglement between blocks of $m$ spins each, one beginning at spin
1 and the other beginning at spin $L$. 
With $A$ and $B$ chosen as in Eqs. (\ref{AB_def}),
calculations similar to the 
derivation of Eqs. (\ref{f20})-(\ref{f21}) show that the entanglement
condition now takes on the form
\begin{eqnarray}
\label{f22}
   &&  Q = \biggl( \sum_{i=1}^m \sum_{j=L+1}^{L+m} I_{1ij} \biggr)^2 
    - \Biggl\{ \biggl[m I_2 + 2\sum_{i=1}^m \sum_{i'=i+1}^{m} I_{3ii'} 
          \biggr]^2 
\nonumber\\
   &&  + \biggl(\sum_{i=1}^m\sum_{j=L+1}^{L+m} I_{1ij} \biggr)
       \biggl(\sum_{i=1}^m\sum_{j=L+1}^{L+m} I_{3ij} \biggr) \Biggr\} >0,
\end{eqnarray}
where $I_{1ij}$  and $I_{3ij}$ are given by the respective 
Eqs. (\ref{f20.1}) and (\ref{f20.3}) with $\Delta r = |{\bf r}_1-{\bf r}_2|$
being replaced by $\Delta r_{ij} = |{\bf r}_i-{\bf r}_j|$. Again
$\langle A^\dagger A B^\dagger B \rangle$ (the term in the curly brackets)
consists of two parts, the first representing the correlations between spins within a block and
the second representing the correlations
between blocks. Equation (\ref{f22}) is general in that the atoms can be
arranged in an arbitrary manner in space. The only assumption used is
that the two blocks do not overlap. As in the case of individual spins,
in the high temperature limit $y_0\rightarrow 0$, 
$I_{3ij} \rightarrow I_{1ij}$, indicating explicitly that the inequality
(\ref{f22}) cannot be satisfied. At low temperatures, whether $Q$ is positive
or negative depends on the details of the terms in the sums over $I_{1ij}$.
\begin{figure}
\includegraphics{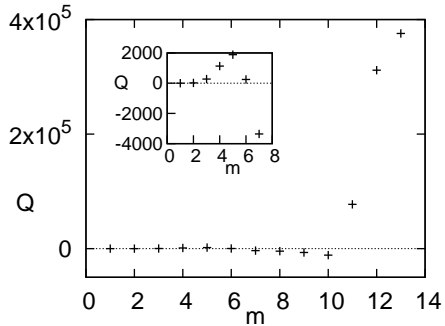}
\caption{
 The quantity $Q$ as a function of the block size $m$ -- the number of spins in each block for a fixed dimensionless temperature 
$\sqrt{2JS/(k_{\rm B}T)}=7$ and for $L=13$. The inset zooms in the part 
of the plot for small $m$.
}
\label{block}
\end{figure}

In Fig. \ref{block} we plotted $Q$ as a function of the block size $m$, $m$
being the number of spins contained in each block. It is assumed that 
each block consists of neighboring spins located along a straight line, one
beginning at spin 1 and the other beginning at spin $L$ so that the blocks
are separated by $L-m$ spins. For the parameters used in Fig. \ref{block},
the case of individual spins $m=1$ exhibits no entanglement 
(cf. Fig. \ref{dx}). As the size of the blocks $m$ increases 
(Fig. \ref{block}, inset), $Q$ acquires positive values indicating a
presence of inter-block entanglement. The change is not monotonic, however. 
As $m$ increases further, the entanglement detected by our condition
can disappear and reappear, being 
particularly strong for $m=11,12,13$. The sign of $Q$ obviously depends on
whether the $I_{1ij}$ add constructively or destructively. An examination  
of inter-block entanglement may thus offer much richer physics than simply
a study of the entanglement between individual spins. 

\section{Conclusions}
We have presented two entanglement conditions for spin systems that allow us to study the 
entanglement between blocks of spins.  Most tests for entanglement in spin systems test for
either complete separability or for entanglement between individual spins, and the results in this
paper complement those.  We have shown that in some cases the conditions involving blocks
of spins can detect entanglement when tests of individual spins cannot.  It was shown that
our entanglement conditions can detect entanglement in intelligent spin states and in states of a
spin chain containing a small number of spin waves.  This latter result was then extended to show
that entanglement in spin waves at finite temperature can also be detected.

\section*{Acknowledgments}
This research was supported by the National Science Foundation under grant PHY-0903660.

\pagebreak

\end{document}